\documentclass[proceedings]{JHEP3}
\usepackage{epsfig,multicol}

\newbox\mybox

\newcommand\fverb{\setbox\mybox=\hbox\bgroup\verb}
\newcommand\fverbdo{\egroup\medskip\noindent\fbox{\unhbox\mybox}\ }
\newcommand\fverbit{\egroup\item[\fbox{\unhbox\mybox}]}
\conference{Rational sequences for the conductance}
\abstract{
We analyse the expression for the conductance of a quantum wire which is
decribed by an integrable quantum field theory.  In the high temperature 
regime we derive a simple formula for the filling fraction. This expression 
involves only the inverse of a matrix which contains the information of the 
asymptotic phases of the scattering matrix and the solutions of the constant 
thermodynamic Bethe ansatz equations. Evaluating these expressions for 
minimal affine Toda field theory we recover several sequences of rational 
numbers, which are multiples of  the 
famous Jain sequence for the filling fraction occurring in the context of the 
fractional quantum Hall effect. For instance we obtain  $\nu= 4 m/(2m +1)$ 
for $A_{4m-1}$-minimal affine Toda field theory. The matrices involved have 
in general non-rational entries and are not part of  previous classification 
schemes based on integral lattices.}

\title{Rational sequences for the conductance in quantum wires from affine
Toda field theories}
\author{O.A.~Castro-Alvaredo and A.~Fring \\
Institut f\"ur Theoretische Physik, Freie Universit\"at Berlin, \\
Arnimallee 14, D-14195 Berlin, Germany \\
E-mail:\email{olalla/fring@physik.fu-berlin.de}}

\input{tcilatex}

\begin{document}

\section{Introduction}

The quantum \cite{Klitz} and in particular the fractional \cite{St} quantum
Hall effect have attracted an enormous amount of attention both from
theorist \cite{Laugh} and experimentalists (for some very recent experiments
see e.g.~\cite{exp}). The key observation is that when subjecting an
electron gas confined to two space dimensions to a strong uniform magnetic
field, the transverse (Hall) conductance takes on preferably certain
characteristic values $G=e^{2}/h\nu $, whereas the longitudinal conductance
vanishes at these plateaux in complete analogy to the classical Hall effect 
\cite{Hall}. The filling fractions $\nu $ are distinct universal, in the
sense that they are independent of the geometry or type of the material,
rational numbers, which can be determined experimentally to an extremely
high precision. Many, but not all, of the experimentally observed filling
fractions are part of Jain's famous sequence (see \cite{Jain} and references
therein) 
\begin{equation}
\nu =\frac{m}{mp\pm 1}\qquad \quad m,p/2=1,2,3,\ldots  \label{Jain}
\end{equation}%
which results as a theoretical prediction from a composite Fermion theory.

In the following we will show that remarkably multiples of these universal
numbers also quantize the conductance of quantum wires when described by
minimal affine Toda field theories (ATFT) \cite{ATFT}. However, no claims
are made here that the systems studied actually correspond to any concrete
description of the real quantum Hall effect. Nonetheless, one may speculate
as there is a well defined way to reduce from a Chern-Simons type theory (an
established description of the quantum Hall effect) to ATFT, see e.g. \cite%
{HS}.

\section{Conductance in the high temperature regime}

Let us briefly recall \cite{OA45} how to compute the conductance $G$ within
the framework of the Landauer-B\"{u}ttiker transport theory \cite{Land} as a
function of the temperature $T$ and elaborate on that expression. Let us
consider a one dimensional quantum wire within the Landauer-B\"{u}ttiker
transport theory. In order to compute $G$ we simply have to determine the
difference of the static charge distribution at the left and right
constriction of the wire, which we assume to be at the potentials $\mu
_{i}^{l}$ and $\mu _{i}^{r}$, respectively. Then, to obtain the direct
current $I_{i}$ for each particle of type $i$ with charge $q_{i}$, we have
to integrate the density distribution functions $\rho _{i}^{r}(\theta ,T,\mu
_{i})$ of occupied states over the full range of the rapidities $\theta $
and the total conductance simply reads 
\begin{eqnarray}
G(1/T) &=&\sum_{i}G_{i}=\sum_{i}\lim_{\Delta \mu _{i}\rightarrow 0}\frac{1}{%
\Delta \mu _{i}}I_{i}(1/T,\Delta \mu _{i}=\mu _{i}^{l}-\mu _{i}^{r}) \\
&=&\sum_{i}\lim_{\Delta \mu _{i}\rightarrow 0}\frac{q_{i}}{2\Delta \mu _{i}}%
\int\limits_{-\infty }^{\infty }d\theta \left[ \rho _{i}^{r}(\theta ,T,\mu
_{i}^{l})-\rho _{i}^{r}(\theta ,T,\mu _{i}^{r})\right] .  \label{1}
\end{eqnarray}%
where $G_{i}$ denotes the contribution to the conductance of each particle $%
i $, and the sums above run both over particles and antiparticles. That
explains the factor of $1/2$ in (\ref{1}) wich accounts for the double
counting. Hence, the main task in this approach is to determine the density
distribution functions $\rho _{i}^{r}(\theta ,T,\mu _{i})$ of occupied
states. It is remarkable that in the context of integrable models, despite
the fact that these functions are neither Fermi-Dirac nor Bose-Einstein,
there exist approaches in which they can be computed non-perturbatively,
i.e.~the thermodynamic Bethe ansatz (TBA) \cite{TBAZam}.

We briefly recall how this is possible. The central equations of the TBA
relate the total density of available states $\rho _{i}(\theta ,r)$ for
particles of type $i$ with mass $m_{i}$ as a function of the inverse
temperature $r=1/T$ to the density of occupied states $\rho _{i}^{r}(\theta
,r)$ 
\begin{equation}
\rho _{i}(\theta ,r)=\frac{m_{i}}{2\pi }\cosh \theta
+\sum\limits_{j}[\varphi _{ij}\ast \rho _{j}^{r}](\theta )\,.  \label{rho}
\end{equation}

\noindent By $\left( f\ast g\right) (\theta )$ $:=1/(2\pi )\int d\theta
^{\prime }f(\theta -\theta ^{\prime })g(\theta ^{\prime })$ we denote as
usual the convolution of two functions. There are only two inputs into the
entire TBA analysis: first\ the dynamical interaction, which enters via the
logarithmic derivative of the scattering matrix $\varphi _{ij}(\theta
)=-id\ln S_{ij}(\theta )/d\theta $ and an assumption on the statistical
interaction $g_{ij}$ amongst the particles $i$ and $j$ on which we comment
further below. For the moment we chose this interaction to be of fermionic
type. The mutual ratio of the two types of densities serves as the
definition of the so-called pseudo-energies $\varepsilon _{i}(\theta ,r)$%
\begin{equation}
\frac{\rho _{i}^{r}(\theta ,r)}{\rho _{i}(\theta ,r)}=\frac{e^{-\varepsilon
_{i}(\theta ,r)}}{1+e^{-\varepsilon _{i}(\theta ,r)}}\,,  \label{dens}
\end{equation}%
which have to be positive and real. At thermodynamic equilibrium they can be
computed from the non-linear integral equations 
\begin{equation}
rm_{i}\cosh \theta =\varepsilon _{i}(\theta ,r,\mu _{i})+r\mu
_{i}+\sum\limits_{j}[\varphi _{ij}\ast \ln (1+e^{-\varepsilon _{j}})](\theta
)\,,  \label{TBA}
\end{equation}%
where $r=m/T$, $m_{l}\rightarrow m_{l}/m$, $\mu _{i}\rightarrow \mu _{i}/m$,
with $m$ being the mass of the lightest particle in the model and chemical
potential $\mu _{i}<1$. As pointed out already in \cite{TBAZam} (here just
with the small modification of a chemical potential), the comparison between
(\ref{TBA}) and (\ref{rho}) leads to the useful relation 
\begin{equation}
\rho _{i}(\theta ,r,\mu _{i})=\frac{1}{2\pi }\left( \frac{d\varepsilon
_{i}(\theta ,r,\mu _{i})}{dr}+\mu _{i}\right) \,\sim \frac{1}{2\pi r}%
\epsilon (\theta )\frac{d\varepsilon _{i}(\theta ,r,\mu _{i})}{d\theta }.
\label{rhoe}
\end{equation}%
Here $\epsilon (\theta )=\Theta (\theta )-\Theta (-\theta )$ is the unit
step function, i.e.~$\epsilon (\theta )=1$ for $\theta >0$ and $\epsilon
(\theta )=-1$ for $\theta <0$. In equation (\ref{dens}), we assume that in
the large rapidity regime the density $\rho _{i}^{r}(\theta ,r,\mu _{i})$ is
dominated by the last expression in (\ref{rhoe}) and in the small rapidity
regime by the Fermi distribution function. Therefore, from (\ref{dens})
follows 
\begin{eqnarray}
\rho _{i}^{r}(\theta ,r,\mu _{i}) &=&\frac{e^{-\varepsilon _{i}(\theta
,r,\mu _{i})}}{1+e^{-\varepsilon _{i}(\theta ,r,\mu _{i})}}\rho _{i}(\theta
,r,\mu _{i}) \\
&\sim &\frac{1}{2\pi r}\epsilon (\theta )\frac{d}{d\theta }\ln \left[ 1+\exp
(-\varepsilon _{i}(\theta ,r,\mu _{i}))\right] \,\,.
\end{eqnarray}%
Using this expression in equation (\ref{1}), we can approximate the direct
current in the ultraviolet by 
\begin{equation}
\lim\limits_{r\rightarrow 0}I_{i}(r,\Delta \mu _{i})\sim \frac{q_{i}}{4\pi r}%
\int\limits_{-\infty }^{\infty }d\theta \ln \left[ \frac{1+\exp
(-\varepsilon _{i}(\theta ,r,\mu _{i}^{l}))}{1+\exp (-\varepsilon
_{i}(\theta ,r,\mu _{i}^{r}))}\right] \,\frac{d\epsilon (\theta )\,}{d\theta 
}\,,  \label{ya}
\end{equation}%
after a partial integration. Taking now the potentials at the end of the
wire to be $\mu _{i}^{r}=-\mu _{i}^{l}=\mu _{i}/2$ we carry out the limit $%
\Delta \mu _{i}\rightarrow 0$ in (\ref{1}) with the help l'H\^{o}pital rule
and the conductance becomes 
\begin{equation}
\lim\limits_{r\rightarrow 0}G_{i}(r)\sim \frac{q_{i}}{2\pi r}%
\int\limits_{-\infty }^{\infty }d\theta \frac{1}{1+\exp [\varepsilon
_{i}(\theta ,r,0)]}\left. \frac{d\varepsilon _{i}(\theta ,r,\mu _{i}/2)}{%
d\mu _{i}}\right\vert _{\mu _{i}=0}\frac{d\epsilon (\theta )\,}{d\theta }\,.
\end{equation}%
Noting that $d\epsilon (\theta )/d\theta =2\delta (\theta )$, we obtain 
\begin{equation}
\lim\limits_{r\rightarrow 0}G_{i}(r)\sim \frac{q_{i}}{\pi r}\frac{1}{1+\exp
\varepsilon _{i}(0,r,0)}\left. \frac{d\varepsilon _{i}(0,r,\mu _{i}/2)}{d\mu
_{i}}\right\vert _{\mu _{i}=0}\,.  \label{g0}
\end{equation}%
The derivative $d\varepsilon _{i}(0,r,\mu _{i}/2)/d\mu _{i}$ can be obtained
by solving 
\begin{equation}
\frac{d\varepsilon _{i}(0,r,\mu _{i}/2)}{d\mu _{k}}=-\frac{r}{2}\delta
_{ik}+\sum_{j}N_{ij}\frac{1}{1+\exp \varepsilon _{j}(0,r,\mu _{i}/2)]}\frac{%
d\varepsilon _{j}(0,r,\mu _{j}/2)}{d\mu _{k}}\,,  \label{de}
\end{equation}%
which results from performing a constant TBA analysis on the $\mu _{k}$%
-derivative of (\ref{TBA}) in the spirit of \cite{TBAZam}. At this point
only the asymptotic phases of the scattering matrix enter via 
\begin{equation}
N_{ij}=\frac{1}{2\pi i}\lim_{\theta \rightarrow \infty }[\ln [S_{ij}(-\theta
)/S_{ij}(\theta )]]~.
\end{equation}%
In principle we have now all quantities needed to compute the conductance,
but to solve (\ref{de}) for the derivatives of the pseudo-energies is
somewhat cumbersome, see \cite{OA45} for such a computation. Nonetheless, we
can elaborate more on equation (\ref{de}) and simplify the procedure
further. For this purpose we introduce the quantity 
\begin{equation}
Y_{ij}:=\frac{1}{r(1+e^{\varepsilon _{i}})}\frac{d\varepsilon _{i}}{d\mu _{j}%
}~,
\end{equation}%
such that we can re-write equation (\ref{de}) equivalently as 
\begin{equation}
M_{ij}Y_{jk}=\frac{\delta _{ik}}{2}\quad \quad \text{with\qquad }%
M_{ij}:=N_{ij}-(1+e^{\varepsilon _{i}})\delta _{ij}  \label{M}
\end{equation}%
where the pseudoenergies satisfy the constant TBA equations 
\begin{equation}
e^{-\varepsilon _{i}}=\prod\limits_{j}(1+e^{-\varepsilon _{j}})^{N_{ij}}~.
\label{ctba}
\end{equation}%
Returning now to dimensionful variables, i.e.~replacing $1/2\pi \rightarrow
e^{2}/h$, the conductance at high temperature in terms of the filling
fraction $\nu $ then simply results to 
\begin{equation}
G(0)=\frac{e^{2}}{h}\nu \quad \quad \text{with\qquad }\nu
=2\sum_{i,j}q_{i}(M^{-1})_{ij}~.  \label{nu}
\end{equation}%
This means we have reduced the entire problem to compute filling fractions
simply to the task of finding and inverting the matrix $M$. This is done in
two steps: First from the asymptotic phases of the scattering matrix we
compute $N_{ij}$ and subsequently we solve the constant TBA equations (\ref%
{ctba}). Then it is a simple matter of inverting the matrix (\ref{M}) and
performing the sums in (\ref{nu}).

In the context of the fractional quantum Hall effect one encounters very
often particles which obey some exotic (anyonic) statistics. So far we have
assumed our particles to obey fermionic type statistics as this choice is
most natural for the investigated theories \cite{TBAZam}. However, one can
easily implement more general statistics by adding a matrix $g_{ij}$ to the $%
N$-matrix \cite{BF}.

The formula (\ref{nu}) reminds of course on the well-known expressions for
the conductance as may be found for instance in \cite{FZ,Capp}. In that
context it was found \cite{FZ,WZ} that Jain's sequence (\ref{Jain}) can be
obtained simply from the $(m\times m)$-matrix 
\begin{equation}
M_{ij}=p\pm \delta _{ij}~.  \label{JJ}
\end{equation}%
For this we have to take $q_{i}=1/2$ $\forall ~i$ in our expression (\ref{nu}%
). We will now demonstrate that a sequence closely related to (\ref{Jain})
can also be obtained in a more surprising way from fairly complicated
matrices, even with non-rational entries, which result directly in the way
indicated above, namely from a TBA analysis of minimal affine Toda field
theories \cite{ATFT}. Each Toda theory is associated to a Lie algebra 
\textbf{g} of rank $\ell $ and it is well known \cite{TBAKM} that in that
case $N$ is an $(\ell \times \ell )$-matrix which is of the general form 
\begin{equation}
N_{ij}=\delta _{ij}-2(K_{\mathbf{g}}^{-1})_{ij}~,  \label{NK}
\end{equation}%
where $K_{\mathbf{g}}$ is the Cartan matrix related to \textbf{g } (see e.g.~%
\cite{Hum}). The solutions to the constant TBA equations are also known \cite%
{Resh,TBAKM} for most cases. In the ultraviolet limit these theories possess
Virasoro central charge $c=2\ell /(H+2)$, with $H$ being the Coxeter number
of the Lie algebra \textbf{g}.

\section{Fractional filling fractions from minimal affine Toda field theory}

\subsection{The $4m/(2m+1)$-sequence}

\noindent\ Let us start with some concrete examples to illustrate the
working of our formulae. Specializing the general expression (\ref{NK}) to
the $A_{3}$-case, the solutions to the constant TBA equations (\ref{ctba})
are simply 
\begin{equation}
e^{\varepsilon _{1}}=e^{\varepsilon _{3}}=2,\qquad e^{\varepsilon _{2}}=3~.
\end{equation}%
Then, the inverse of the $M$-matrix 
\begin{equation}
M_{ij}=\delta _{ij}-2(K_{A_{3}}^{-1})_{ij}-\delta _{ij}(1+e^{\varepsilon
_{i}})~
\end{equation}%
is computed to 
\begin{equation}
M^{-1}=\frac{1}{36}\left( 
\begin{array}{ccc}
11 & -2 & -1 \\ 
-2 & 8 & -2 \\ 
-1 & -2 & 11%
\end{array}%
\right) ~.
\end{equation}%
From the fact that the $A_{\ell }$-minimal affine Toda field theories can
also be viewed as complex sine-Gordon models \cite{CSG}, we know \cite{DH}
that the charges in this theory are $q_{1}=q_{3}=1$, $q_{2}=2$, such that (%
\ref{nu}) yields 
\begin{equation}
\nu _{A_{3}}=4/3~.
\end{equation}
The next example, i.e.~$A_{5}$-minimal affine Toda field theory, yields a
less expected answer, even more since the $M$-matrix contains non-rational
entries. With (\ref{NK}) for $A_{5}$ the solutions to the constant TBA
equations are \cite{Resh,TBAKM} 
\begin{equation}
e^{\varepsilon _{1}}=e^{\varepsilon _{5}}=1+\sqrt{2},\qquad e^{\varepsilon
_{2}}=e^{\varepsilon _{4}}=2+2\sqrt{2},\quad \quad e^{\varepsilon _{3}}=3+2%
\sqrt{2}~.
\end{equation}%
Assembling this into the $M$-matrix, it is clear that it will contain
non-rational entries. Evidently this matrix is not of the \ form (\ref{JJ})
and certainly falls out of the classification scheme based on integral
lattices \cite{FT}. Nonetheless, it will lead to a distinct rational value
for $\nu $. We compute the inverse of $M$ to 
\begin{equation}
M^{-1}=\left( 
\begin{array}{ccccc}
\left( \frac{35}{4}-6\sqrt{2}\right) & \left( {\frac{31}{2\,\sqrt{2}}}%
-11\right) & \left( {\frac{7-5\,\sqrt{2}}{4}}\right) & \left( 6-{\frac{17}{%
2\,\sqrt{2}}}\right) & \left( 3\,\sqrt{2}-{\frac{17}{4}}\right) \\ 
\left( {\frac{31}{2\,\sqrt{2}}}-11\right) & \left( 15-{\frac{21}{\sqrt{2}}}%
\right) & \left( {\frac{7\,\sqrt{2}-10}{4}}\right) & \left( 6\,\sqrt{2}-{%
\frac{17}{2}}\right) & \left( 6-{\frac{17}{2\,\sqrt{2}}}\right) \\ 
\left( {\frac{7-5\,\sqrt{2}}{4}}\right) & \left( {\frac{7\,\sqrt{2}-10}{4}}%
\right) & \left( {\frac{9}{4}}-{\frac{3}{\sqrt{2}}}\right) & \left( {\frac{%
7\,\sqrt{2}-10}{4}}\right) & \left( {\frac{7-5\,\sqrt{2}}{4}}\right) \\ 
\left( 6-{\frac{17}{2\,\sqrt{2}}}\right) & \left( 6\,\sqrt{2}-{\frac{17}{2}}%
\right) & \left( {\frac{7\,\sqrt{2}-10}{4}}\right) & \left( 15-{\frac{21}{%
\sqrt{2}}}\right) & \left( {\frac{31}{2\,\sqrt{2}}}-11\right) \\ 
\left( 3\,\sqrt{2}-{\frac{17}{4}}\right) & \left( 6-{\frac{17}{2\,\sqrt{2}}}%
\right) & \left( {\frac{7-5\,\sqrt{2}}{4}}\right) & \left( {\frac{31}{2\,%
\sqrt{2}}}-11\right) & \left( {\frac{35}{4}}-6\,\sqrt{2}\right)%
\end{array}%
\right) ~.  \label{M5}
\end{equation}%
Remarkably when taking into account that \cite{DH} $q_{1}=q_{5}=1$, $%
q_{2}=q_{4}=2$, $q_{3}=3$, we obtain by evaluating (\ref{nu}) for the matrix
(\ref{M5}) the simple ratio 
\begin{equation}
\nu _{A_{5}}=3/2~.
\end{equation}%
We will now turn to the generic case. Taking the general solutions of the
constant TBA equations into account \cite{Resh,TBAKM} and using a generic
expression for the inverse of the Cartan matrix $K_{A_{\ell }}^{-1}=\min
(i,j)-ij/(\ell +1)$ in (\ref{NK}), the M-matrix for an $A_{2\ell +1}$%
-minimal affine Toda field theory can be written generically as 
\begin{equation}
M_{ij}=\frac{ij}{\ell +1}-2\min (i,j)-\delta _{ij}\frac{\sin \left( \frac{%
i\pi }{2\ell +4}\right) \sin \left( \frac{(i+2)\pi }{2\ell +4}\right) }{\sin
^{2}\left( \frac{\pi }{2\ell +4}\right) }~.  \label{Ml}
\end{equation}%
As already indicated by the previous example this matrix is not of the form (%
\ref{JJ}) and does not fit into the classification scheme proposed in \cite%
{FT}. According to \cite{DH} we have the charges 
\begin{equation}
q_{i}=q_{2\ell +2-i}\qquad \text{and\qquad }q_{i}=i\quad \text{for }i\leq
\ell +1~.  \label{ql}
\end{equation}%
As can be guessed from (\ref{M5}), it is not evident how to express the
inverse in terms of a simple closed expression. We can, however, invert (\ref%
{Ml}) case-by-case up to very high rank and we obtain from (\ref{nu})
together with (\ref{ql}) the sequence 
\begin{equation}
\nu _{A_{2\ell +1}}=\frac{2\ell +2}{\ell +2}~.  \label{JP22}
\end{equation}%
In view of (\ref{Ml}), it is remarkable that the outcome is rational. Note
for $\ell =0$, that is $A_{1}$ we recover the free case with $\nu =1$.
Taking now $\ell =2m-1$, we obtain as a subsequence of this four times the
most stable part of Jain's sequence (\ref{Jain}) with $p=2$%
\begin{equation}
\nu _{A_{4m-1}}=\frac{4m}{2m+1}~.  \label{Jp2}
\end{equation}%
In summary: \emph{The conductance of a quantum wire which is described by a
massive }$A_{2\ell +1}$\emph{-minimal affine Toda field theory possesses in
the high temperature regime, in which the model turns into a conformal field
theory with Virasoro central }$c=(2\ell +1)/(\ell +2)$\emph{, a filling
fraction equal to (\ref{JP22}). In particular for }$\ell =2m-1$\emph{, we
obtain the sequence (\ref{Jp2}).}

\subsection{The $2m/(2m+1)$-sequence}

\noindent We proceed now similarly as in the preceding section, but now for
the $D_{2\ell +1}$-minimal affine Toda field theories, which all possess
Virasoro central charge $c=1$ in the ultraviolet limit. We label the
particles in consecutive order along the Dynkin diagram (see e.g.~\cite{Hum}
for more properties), starting from the not splitted end. Taking in (\ref{NK}%
) \textbf{g}=$D_{2\ell +1}$ the solutions to the constant TBA equations are
simply \cite{Resh,TBAKM} 
\begin{eqnarray}
e^{\varepsilon _{i}} &=&i(i+2)\qquad 1\leq i\leq 2\ell -1  \label{T1} \\
e^{\varepsilon _{2\ell +1}} &=&e^{\varepsilon _{2\ell }}=2\ell ~.  \label{T2}
\end{eqnarray}%
Since these entries are all integer valued, we are not very surprised when
we obtain rational values for the filling fraction, but what is not obvious
is that the outcome is one of Jain's sequences. The $M$-matrix is computed
to 
\begin{equation}
M_{ij}=-2(K_{D_{2\ell +1}}^{-1})_{ij}-\delta _{ij}e^{\varepsilon _{i}}~,
\label{MD}
\end{equation}%
with the values (\ref{T1}) and (\ref{T2}). From these data we evaluate a
simple expression for the determinant 
\begin{equation}
\det M=\frac{(2\ell +1)^{2}~(2\ell +1)!~(2\ell )!}{2}~,
\end{equation}%
and the inverse of this matrix 
\begin{eqnarray}
(M^{-1})_{ij} &=&(M^{-1})_{ji}=\frac{2}{3j(1+j)(2+j)}\qquad ~~2\leq i<j\leq
2\ell -1 \\
(M^{-1})_{ii} &=&\frac{-(3i+1)}{3i(1+i)(2+i)}\qquad \qquad \qquad \qquad
1\leq i\leq 2\ell -1 \\
(M^{-1})_{i(2\ell +1)} &=&(M^{-1})_{i(2\ell )}=\frac{1}{6\ell (2\ell +1)}%
\qquad \qquad ~1\leq i\leq 2\ell -1 \\
(M^{-1})_{(2\ell +1)i} &=&(M^{-1})_{(2\ell )i}=\frac{1}{6\ell (2\ell +1)}%
\qquad \qquad ~1\leq i\leq 2\ell -1 \\
(M^{-1})_{(2\ell +1)(2\ell +1)} &=&(M^{-1})_{(2\ell )(2\ell )}=-\frac{10\ell
+1}{12\ell (2\ell +1)} \\
(M^{-1})_{(2\ell +1)(2\ell )} &=&(M^{-1})_{(2\ell )(2\ell +1)}=\frac{2\ell -1%
}{12\ell (2\ell +1)}
\end{eqnarray}%
Taking then the charges of the particles to be 
\begin{equation}
q_{2\ell +1}=q_{2\ell }=\ell /2\qquad \text{and\qquad }q_{i}=i\quad \text{%
for }i\leq 2\ell -1~,
\end{equation}%
the computation of (\ref{nu}) yields 
\begin{equation}
\nu _{D_{2\ell +1}}=\frac{2\ell }{2\ell +1}~.  \label{JP33}
\end{equation}%
Similarly as in the previous subsection, the sequence (\ref{JP33})\emph{\ }%
gives twice the Jain sequence (\ref{Jain}) with $p=2.$

In summary: \emph{The conductance of a quantum wire which is described by a
massive }$D_{2\ell +1}$\emph{-minimal affine Toda field theory possesses in
the high temperature regime, in which all models turn into conformal field
theories with Virasoro central charge }$c=1$\emph{, a filling fraction equal
to (\ref{JP33}) which is twice the principal Jain sequence (\ref{Jain}).}

\subsection{The $4m/(6m+1)$-sequence}

\noindent This sequence can be obtained similarly just by altering the
values of the two charges at the very end of the Dynkin diagram. Considering
now the $D_{6m+2}$-minimal affine Toda field theories we can employ the same 
$M$-matrix as in the previous subsection, but we take the charges of the
particles to be 
\begin{equation}
q_{6m+2}=q_{6m+1}=m/(2m+1)\qquad \text{and\qquad }q_{i}=i\quad \text{for }%
i\leq 6m~.
\end{equation}%
Evaluating then the expression for the filling fractions (\ref{nu}) gives 
\begin{equation}
\nu _{D_{6m+2}}=\frac{4m}{6m+1}~,
\end{equation}%
which is four times the Jain's sequence (\ref{Jain}) with $p=6$.

\section{Conclusions}

Within a Landauer-B\"{u}ttiker transport theory picture we have analyzed the
expression for the conductance of a quantum wire which is described by an
integrable quantum field theory. The final expression for the conductance in
the high temperature regime is very simple (\ref{nu}) and involves the sum
over the entries of the inverse of a certain matrix $M$ as defined in (\ref%
{M}). This matrix is constructed from the knowledge of the asymptotic phases
of the scattering matrix and the solutions of the constant TBA equations (%
\ref{ctba}).

When evaluating this matrix for some concrete minimal affine Toda field
theories, we obtain values for the filling fraction which coincide with
multiples of several subsequences of Jain's series (\ref{Jain}) and are
therefore rational numbers. The fact that we obtain this special rational
values is extremely surprising, in particular as for the $A_{2\ell +1}$%
-minimal affine Toda theories the related $M$-matrix has non-rational
entries. One should note, however, that one does not always get these nice
rational values. We did not report all examples here which we have computed,
but for instance in general the $A_{2\ell }$ and the $D_{2\ell }$-minimal
affine Toda theories lead to non-rational values for $\nu $.

Our findings pose several interesting questions: As it is clear that the $M$%
-matrices obtained are beyond the classification scheme carried out in \cite%
{FT} on the basis of integral lattices, one may attempt a new type of
classification based on the Lie algebraic systematics which underlies the
formulation of integrable quantum field theories. In order to do this we
have to enlarge our considerations \cite{OA55} to other algebras such as the 
$E$-series, non-simply laced Lie algebras and also to theories which are
related to a pair of Lie algebras. It would also be interesting to perform
an analysis based on a different expression from (\ref{1}) for the
conductance, such as the Kubo formula, and compare the findings similar as
in \cite{OA45}.

\noindent \textbf{Acknowledgments: }We are grateful to the Deutsche
Forschungsgemeinschaft (Sfb288), for financial support.

\end{document}